\documentclass[runningheads]{llncs}
\usepackage{bm}
\usepackage{amsfonts}
\usepackage{srcltx}
\usepackage{graphicx}
\usepackage{caption}
\usepackage{subcaption}
\usepackage{amsmath}

\usepackage{amsthm}
\usepackage{mathtools}
\usepackage{microtype}
\usepackage{enumerate}
\usepackage{xcolor}
\usepackage{here}
\usepackage{authblk}
\usepackage{lineno}
\usepackage{booktabs}   
\usepackage{caption}
\usepackage{subcaption}
\usepackage{multirow}
\usepackage{booktabs}
\usepackage{makecell}
\usepackage{cite}
\usepackage[misc]{ifsym}
\usepackage{lipsum}

\begin{document}
\title{Diffusion-based Data Augmentation for Nuclei Image Segmentation}
%
%
\author{
Xinyi Yu\inst{1}
\and 
Guanbin Li\inst{2}
\and 
Wei Lou\inst{3}
\and 
Siqi Liu\inst{1}
\and 
Xiang Wan\inst{1}
\and 
Yan Chen\inst{4}
\and 
Haofeng Li\inst{1}$^{(\textrm{\Letter})}$
}

%
\authorrunning{X. Yu et al.}
%

\institute{Shenzhen Research Institute of Big Data, Shenzhen, China 
\and 
School of Computer Science and Engineering, Research Institute of Sun Yat-sen University in Shenzhen, Sun Yat-sen University, Guangzhou, China 
\and 
The Chinese University of Hong Kong, Shenzhen, China 
\and 
Shenzhen Health Development Research and Data Management Center\\
\email{lhaof@sribd.cn}
}

\maketitle              

\newcommand\blfootnote[1]{%
\begingroup
\renewcommand\thefootnote{}\footnote{#1}%
\addtocounter{footnote}{-4}%
\endgroup
}

\begin{abstract}
Nuclei segmentation is a fundamental but challenging task in the quantitative analysis of histopathology images. Although fully-supervised deep learning-based methods have made significant progress, a large number of labeled images are required to achieve great segmentation performance. Considering that manually labeling all nuclei instances for a dataset is inefficient, obtaining a large-scale human-annotated dataset is time-consuming and labor-intensive.  
Therefore, augmenting a dataset with only a few labeled images to improve the segmentation performance is of significant research and application value. In this paper, we introduce the first diffusion-based augmentation method for nuclei segmentation. The idea is to synthesize a large number of labeled images to facilitate training the segmentation model. To achieve this, we propose a two-step strategy. In the first step, we train an unconditional diffusion model to synthesize the \textit{Nuclei Structure} that is defined as the representation of pixel-level semantic and distance transform. Each synthetic nuclei structure will serve as a constraint on histopathology image synthesis and is further post-processed to be an instance map. In the second step, we train a conditioned diffusion model to synthesize histopathology images based on nuclei structures. The synthetic histopathology images paired with synthetic instance maps will be added to the real dataset for training the segmentation model. The experimental results show that by augmenting 10\% labeled real dataset with synthetic samples, one can achieve comparable segmentation results with the fully-supervised baseline. The code is released in: https://github.com/lhaof/Nudiff 
\blfootnote{This work is supported by Chinese Key-Area Research and Development Program of Guangdong Province (2020B0101350001), 
and the Guangdong Basic and Applied Basic Research Foundation (2023A1515011464, 2020B1515020048), 
and the National Natural Science Foundation of China (No.~62102267, No.~61976250), 
and the Shenzhen Science and Technology Program (JCYJ20220818103001002, JCYJ20220530141211024), 
and the Guangdong Provincial Key Laboratory of Big Data Computing, The Chinese University of Hong Kong, Shenzhen. 
%
Haofeng Li is the corresponding author.
}
\keywords{Data augmentation \and Nuclei segmentation \and Diffusion models.}
\end{abstract}
\section{Introduction}
Nuclei segmentation is a fundamental step in medical image analysis. Accurately segmenting nuclei helps analyze histopathology images to facilitate clinical diagnosis and prognosis. In recent years, many deep learning based nuclei segmentation methods have been proposed~\cite{naylor2018segmentation,graham2019hover,liu2021panoptic,liu2020unsupervised}. Most of these methods are fully-supervised so the great segmentation performance usually relies on a large number of labeled images. However, manually labeling the pixels belonging to all nucleus boundaries in an image is time-consuming and requires domain knowledge. In practice, it is hard to obtain an amount of histopathology images with dense pixel-wise annotations but feasible to collect a few labeled images.
A question is raised naturally: can we expand the training dataset with a small proportion of images labeled to reach or even exceed the segmentation performance of the fully-supervised baseline? Intuitively, since the labeled images are samples from the population of histopathology images, if the underlying distribution of histopathology images is learned, one can generate infinite images and their pixel-level labels to augment the original dataset. Therefore, it is demanded to develop a tool that is capable of learning distributions and generating new paired samples for segmentation. 

\begin{figure}[!t]
\centering
\includegraphics[width=1.\textwidth]{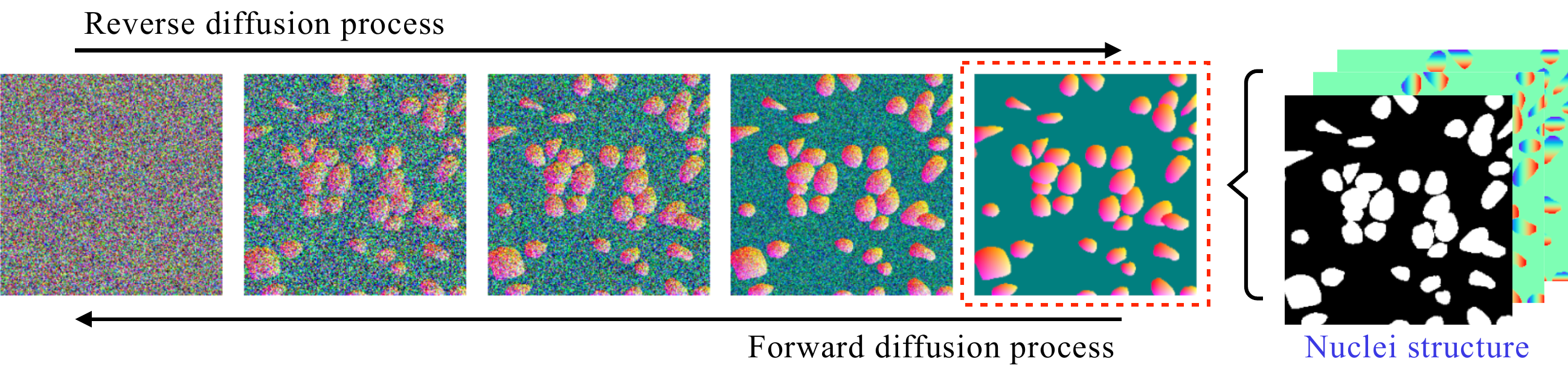}
\caption{The illustration of diffusion model in the context of nuclei structure.} 
\label{ddpm-demo}
\end{figure}

Generative adversarial network (GANs) \cite{goodfellow2014generative,arjovsky2017wasserstein,li2018context,karras2019style,lou2022pixel} have been widely used in data augmentation \cite{mirza2014conditional,isola2017image,zhu2017unpaired,shaham2019singan}. Specially, a newly proposed GAN-based method can synthesize labeled histopathology image for nuclei segmentation \cite{lou2023multi}. While GANs are able to generate high quality images, they are known for unstable training and lack of diversity in generation due to the adversarial training strategy. Recently, diffusion models represented by denoising diffusion probabilistic model (DDPM) \cite{ho2020denoising} tend to overshadow GANs. Due to the theoretical basis and impressive performance of diffusion models, they were soon applied to a variety of vision tasks, such as inpainting, superresolution~\cite{yue2021robust}, text-to-image translation, anomaly detection and segmentation \cite{rombach2022high,nichol2022glide,ho2022cascaded,amit2021segdiff}. As likelihood-based models, diffusion models do not require adversarial training and outperform GANs on the diversity of generated images \cite{dhariwal2021diffusion}, which are naturally more suitable for data augmentation. In this paper, we propose a novel diffusion-based augmentation framework for nuclei segmentation. The proposed method consists of two steps: unconditional nuclei structure synthesis and conditional histopathology image synthesis. We develop an unconditional diffusion model and a nuclei-structure conditioned diffusion model (Fig. \ref{ddpm-demo}) for the first and second step, respectively. On the training stage, we train the unconditional diffusion model using nuclei structures calculated from instance maps and the conditional diffusion model using paired images and nuclei structures. On the testing stage, the nuclei structures and the corresponding images are generated successively by the two models. As far as our knowledge, we are the first to apply diffusion models on histopathology image augmentation for nuclei segmentation. 

Our contributions are: (1) a diffusion-based data augmentation framework that can generate histopathology images and their segmentation labels from scratch; (2) an unconditional nuclei structure synthesis model and a conditional histopathology image synthesis model; (3) experiments show that with our method, by augmenting only 10\% labeled training data, one can obtain segmentation results comparable to the fully-supervised baseline.

\section{Method}
\begin{figure}[t]
\includegraphics[width=\textwidth]{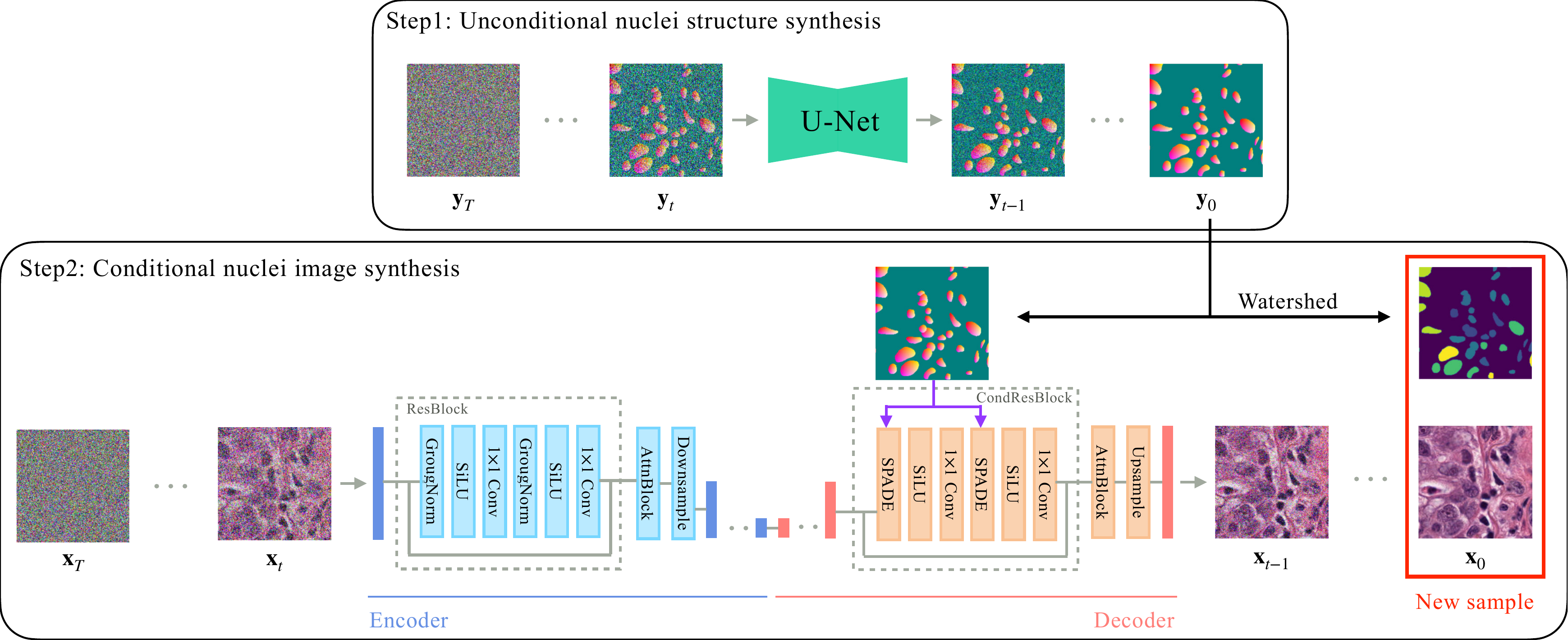}
\caption{The proposed diffusion-based data augmentation framework. We first generate a nuclei structure with an unconditional diffusion model, and then generate images conditioned on the nuclei structure. The instance map from the nuclei structure is paired with the synthetic image to forms a new sample. 
} 
\label{framework}
\end{figure}

Our goal is to augment a dataset containing a limited number of labeled images with more samples to improve the segmentation performance. To increase the diversity of labeled images, it is preferred to synthesize both images and their corresponding instance maps. We propose a two-step strategy for generating new labeled images. Both steps are based on diffusion models. The overview of the proposed framework is shown in Fig. \ref{framework}. In this section, we introduce the two steps in detail.

\subsection{Unconditional Nuclei Structure Synthesis}
In the first step, we aim to synthesize more instance maps. Since it is not viable to directly generate an instance map, we instead choose to generate its surrogate -- \textit{nuclei structure}, which is defined as the concatenation of pixel-level semantic and distance transform. Pixel-level semantic is a binary map where 1 or 0 indicates whether a pixel belongs to a nucleus or not. The distance transform consists of the horizontal and the vertical distance transform, which are obtained by calculating the normalized distance of each pixel in a nucleus to the horizontal and the vertical line passing through the nucleus center \cite{graham2019hover}. Clearly, the nuclei structure is a 3-channel map with the same size as the image. As nuclei instances can be identified from the nuclei structure, we can easily construct the corresponding instance map by performance marker-controlled watershed algorithm on the nuclei structure \cite{yang2006nuclei}. Therefore, the problem of synthesizing instance map transfers to synthesizing nuclei structures. We deploy an unconditional diffusion model to learn the distribution of nuclei structures.

Denote a true nuclei structure as $\mathbf{y}_0$, which is sampled from real distribution $q(\mathbf{y})$. To maximize data likelihood, the diffusion model defines a forward and a reverse process. In the forward process, small amount of Gaussian noise are successively added to the sample $\mathbf{y}_0$ in $T$ steps by:  
\begin{equation}
    \mathbf{y}_t=\sqrt{1-\beta_t}\mathbf{y}_{t-1}+\sqrt{\beta_t}\epsilon_{t-1},t=1,...,T,
\end{equation}
where $\epsilon_t\sim\mathcal{N}(0,\mathbf{I})$ and $\{\beta_t\in(0,1)\}_{t=1}^T$ is a variance schedule. The resulting sequence $\{\mathbf{y}_0,...,\mathbf{y}_T\}$ forms a Markov chain. The conditional probability of $\mathbf{y}_t$ given $\mathbf{y}_{t-1}$ follows a Gaussian distribution:
\begin{equation}
    q(\mathbf{y}_t|\mathbf{y}_{t-1})=\mathcal{N}(\mathbf{y}_t;\sqrt{1-\beta_t}\mathbf{y}_{t-1},\beta_t\mathbf{I}).
\end{equation} 

In the reverse process, since $q(\mathbf{y}_{t-1}|\mathbf{y}_t)$ cannot be easily estimated, a model $p_{\theta}(\mathbf{y}_{t-1}|\mathbf{y}_t)$ (typically a neural network) will be learned to approximate $q(\mathbf{y}_{t-1}|\mathbf{y}_t)$. Specifically, $p_{\theta}(\mathbf{y}_{t-1}|\mathbf{y}_t)$ is a also Gaussian distribution:
\begin{equation}
    p_{\theta}(\mathbf{y}_{t-1}|\mathbf{y}_t)=\mathcal{N}(\mathbf{y}_{t-1};\mathbf{\mu}_{\theta}(\mathbf{y}_t,t),\mathbf{\Sigma}_{\theta}(\mathbf{y}_t,t)),
\end{equation} 

The objective function is the variational lower bound loss: $L=L_T+L_{T-1}+...+L_0$, where every term except $L_0$ is a KL divergence between two Gaussian distributions. In practice, a simplified version of $L_t$ is commonly used \cite{ho2020denoising}:
\begin{equation}
L_t^{simple}=\mathbb{E}_{\mathbf{y}_0,\mathbf{\epsilon}_t}\Vert\mathbf{\epsilon}_t-\mathbf{\epsilon}_\theta(\sqrt{\bar{\alpha}_t}\mathbf{y}_t+\sqrt{1-\bar{\alpha}_t}\mathbf{\epsilon}_t,t)\Vert^2,
\end{equation}
where $\alpha_t=1-\beta_t$ and $\bar{\alpha}_t=\prod_{i=1}^t\alpha_i$. Clearly, the optimization objective of the neural network parameterized by $\theta$ is to predict the Gaussian noise $\mathbf{\epsilon}_t$ from the input $\mathbf{y}_t$ at time t. 

After the network is trained, one can progressively denoise a random point from $\mathcal{N}(0,\mathbf{I})$ by $T$ steps to produce a new sample:
\begin{equation}
    \mathbf{y}_{t-1}=\frac{1}{\sqrt{\alpha_t}}(\mathbf{y}_t-\frac{1-\alpha_t}{\sqrt{1-\bar{\alpha}_t}}\mathbf{\epsilon}_\theta(\mathbf{y}_t,t))+\sigma_t\mathbf{z},\,\mathbf{z}\sim\mathcal{N}(0,\mathbf{I})
\end{equation}

For synthesizing nuclei structures, we train an unconditional DDPM on nuclei structures calculated from real instance maps. Following \cite{ho2020denoising}, the network of this unconditional DDPM has a U-Net architecture. 

\subsection{Conditional Histopathology Image Synthesis}
In the second step, we synthesize histopathology images conditioned on nuclei structures. Without any constraint, an unconditional diffusion model will generate diverse samples. There are usually two ways to synthesize images constrained by certain conditions: classifier-guided diffusion \cite{dhariwal2021diffusion} and classifier-free guidance \cite{ho2021classifier}. Since classifier-guided diffusion requires training a separate classifier which is an extra cost, we choose classifier-free guidance to control sampling process. 

Let $\mathbf{\epsilon}_\theta(\mathbf{x}_t,t)$ and $\mathbf{\epsilon}_\theta(\mathbf{x}_t,t,\mathbf{y})$ be the noise predictor of unconditional diffusion model $p_\theta(\mathbf{x}|\mathbf{y})$ and conditional diffusion model $p_\theta(\mathbf{x})$, respectively. The two models can be learned with one neural network. Specifically, $p_\theta(\mathbf{x}|y)$ is trained on paired data $(\mathbf{x}_0,\mathbf{y}_0)$ and $p_\theta(\mathbf{x})$ can be trained by randomly discarding $y$ (i.e $\mathbf{y}=\emptyset$) with a certain $\textit{drop\_rate}\in(0,1)$ so that the model learns unconditional and conditional generation simultaneously. The noise predictor $\mathbf{\epsilon}'_\theta(\mathbf{x}_t,t,y)$ of classifier-free guidance is a combination of the above two predictors:
\begin{equation}
    \mathbf{\epsilon}'_\theta(\mathbf{x}_t,t,y)=(w+1)\mathbf{\epsilon}_\theta(\mathbf{x}_t,t,y)-w\mathbf{\epsilon}_\theta(\mathbf{x}_t,t),
\end{equation}
where $\mathbf{\epsilon}_\theta(\mathbf{x}_t,t)=\mathbf{\epsilon}_\theta(\mathbf{x}_t,t,\mathbf{y}=\emptyset)$, $w$ is a scalar controlling the strength of classifier-free guidance.

Unlike the network of unconditional nuclei structure synthesis which inputs the noisy nuclei structure $\mathbf{y}_t$ and outputs the prediction of $\mathbf{\epsilon}_t(\mathbf{y}_t,t)$, the network of conditional nuclei image synthesis takes the noisy nuclei image $\mathbf{x}_t$ and the corresponding nuclei structure $\mathbf{y}$ as inputs and the prediction of $\mathbf{\epsilon}_t(\mathbf{x}_t,t,\mathbf{y})$ as output. Therefore, the conditional network should be equipped with the ability to well align the paired histopathology image and nuclei structure. Since nuclei structures and histopathology images have different feature spaces, simply concatenating or passing them through a cross-attention module~\cite{li2019motion,he2019non,li2020depthwise} before entering the U-Net will degrade image fidelity and yield unclear correspondence between synthetic nuclei image and its nuclei structure. Inspired by \cite{wang2022semantic}, we embed information of the nuclei structure into feature maps of nuclei image by the spatially-adaptive normalization (SPADE) module \cite{park2019semantic}. In other words, the spatial and morphological information of nuclei modulates the normalized feature maps such that the nuclei are generated in the right places while the background is left to be created freely. We include the SPADE module in different levels of the network to utilize the multi-scale information of nuclei structure. The network of conditional nuclei image synthesis also applies a U-Net architecture. The encoder is a stack of Resblocks and attention blocks (AttnBlocks). Each Resblock consists of 2 GroupNorm-SiLU-Conv and each Attnblocks calculates the self-attention of the input feature map. The decoder is a stack of CondResBlocks and attention blocks. Each CondResBlock consists of SPADE-SiLU-Conv which takes both feature map and nuclei structure as inputs. 

\section{Experiments and Results}
\subsection{Implementation details}
\textbf{Datasets.} We conduct experiments on two datasets: MoNuSeg \cite{kumar2019multi} and Kumar \cite{kumar2017dataset}. The MoNuSeg dataset has 44 labeled images of size $1000\times 1000$, 30 for training and 14 for testing. The Kumar dataset consists of 30 $1000\times 1000$ labeled images from seven organs of The Cancer Genome Atlas (TCGA) database. The dataset is splited into 16 training images and 14 testing images. 

\noindent\textbf{Paired sample synthesis.} To validate the effectiveness of the proposed augmentation method, we create 4 subsets of each training dataset with 10\%, 20\%, 50\% and 100\% nuclei instance labels. Precisely, we first crop all images of each dataset into $256\times 256$ patches with stride 128, then obtain the features of all patches with pretrained ResNet50 \cite{he2016deep} and cluster the patches into 6 classes by K-means. Patches close to the cluster centers are selected. 
The encoder and decoder of the two networks have 6 layers with channels 256, 256, 512, 512, 1024 and 1024. For the unconditional nuclei structure synthesis network, each layer of the encoder and decoder has 2 ResBlocks and last 3 layers contain AttnBlocks. The network is trained using the AdamW optimizer with a learning rate of $10^{-4}$ and a batch size of 4. For the conditional histopathology image synthesis network, each layer of the encoder and the decoder has 2 ResBlocks and 2 CondResBlocks respectively, and last 3 layers contain AttnBlocks. The network is first trained in a fully-conditional style ($\textit{drop\_rate}=0$) and then finetuned in a classifier free style ($\textit{drop\_rate}=0.2$). We use AdamW optimizer with learning rates of $10^{-4}$ and $2\times 10^{-5}$ for the two training stages, respectively. The batch size is set to be 1. For the diffusion process of both steps, we set the total diffusion timestep $T$ to 1000 with a linear variance schedule $\{\beta_1,...,\beta_T\}$ following \cite{ho2020denoising}. 

For MoNuSeg dataset, we generate 512/512/512/1024 synthetic samples for 10\%/20\%/50\%/100\% labeled subsets; for Kumar dataset, 256/256/256/512 synthetic samples are generated for 10\%/20\%/50\%/100\% labeled subsets. The synthetic nuclei structures are generate by the nuclei structure synthesis network and the corresponding images are generated by the histopathology image synthesis network with the classifier-free guidance scale $w=2$. Each follows the reverse diffusion process with 1000 timesteps \cite{ho2020denoising}. We then obtain the augmented subsets by adding the synthetic paired images to the corresponding labeled subsets. 

\noindent\textbf{Nuclei segmentation.} The effectiveness of the proposed augmentation method can be evaluated by comparing the segmentation performance of using the four labeled subsets and using the corresponding augmented subsets to train a segmentation model. We choose to train two nuclei segmentation models -- Hover-Net \cite{graham2019hover} and PFF-Net \cite{liu2021panoptic}. To quantify the segmentation performance, we use two metrics: Dice coefficient and Aggregated Jaccard Index (AJI) \cite{kumar2017dataset}. 

\begin{figure}[htbp]
\centering
\includegraphics[width=\textwidth]{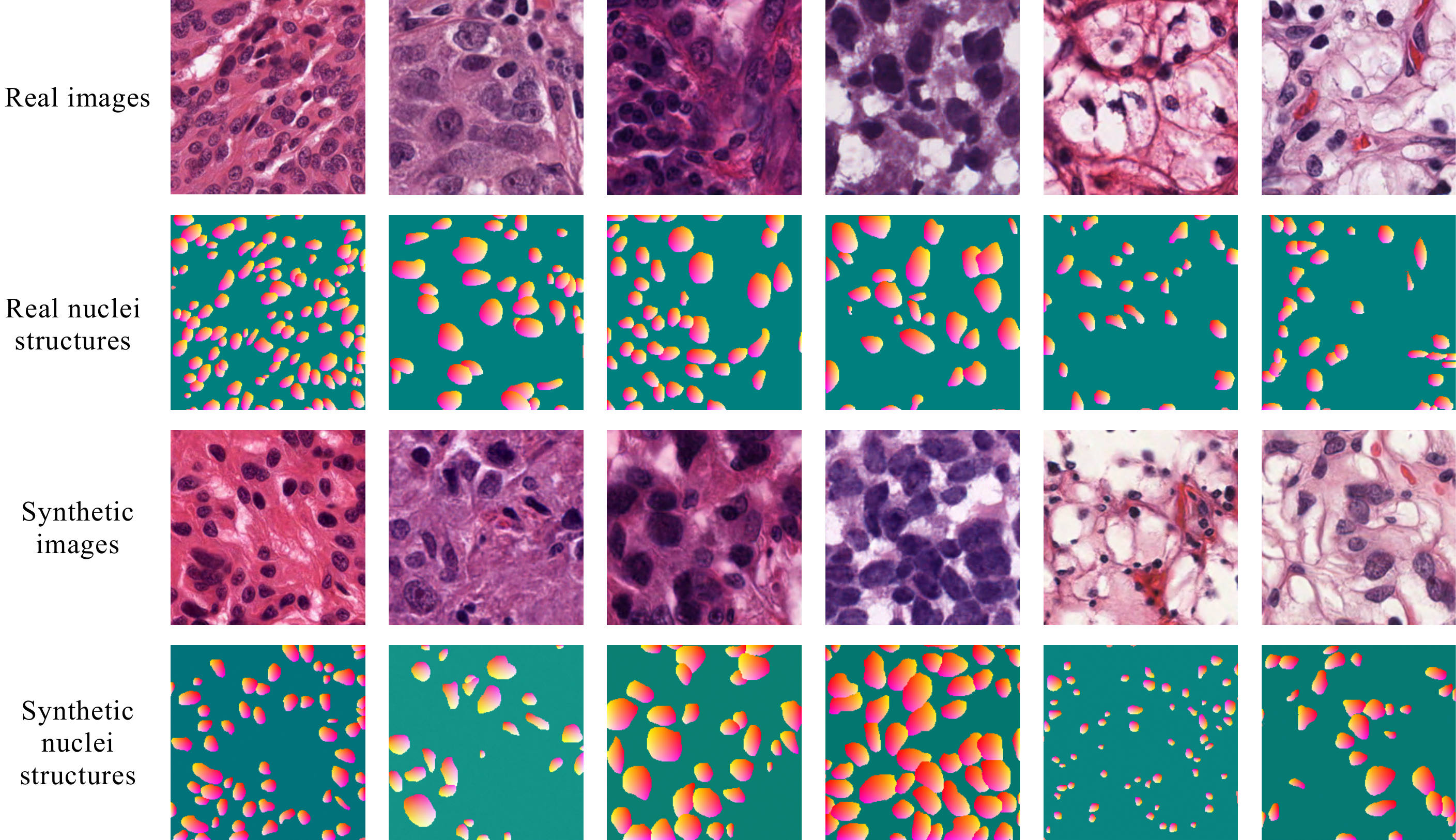}
\caption{Visualization of synthetic samples. The first and second row show selected patches and corresponding nuclei structures from the 10\% labeled subset of MoNuSeg dataset. The third and fourth row show selected synthetic images and corresponding nuclei with similar style as the real one in the same column.} 
\label{viz}
\end{figure}

\subsection{Effectiveness of the proposed data augmentation method}
Fig. \ref{viz} shows the synthetic samples from the models trained on the subset with 10\% labeled images. We have the following observations. First, the synthetic samples look realistic: the patterns of synthetic nuclei structures and textures of synthetic images are close to the real samples. Second, due to the conditional mechanism of the image synthesis network and the classifier-guidance sampling, the synthetic images are well aligned with the corresponding nuclei structures, which is the prerequisite to be additional segmentation training samples. Third, the synthetic nuclei structures and images show great diversity: the synthetic samples resemble different styles of the real ones but with apparent differences. 

We then train segmentation models on the four labeled subsets of MoNuSeg and Kumar dataset and corresponding augmented subsets with both real and synthetic labeled images. With a specific labeling proportion, say 10\%, we name the original subset as 10\% labeled subset and the augmented on as 10\% augmented subset. Specially, 100\% labeled subset is the fully-supervised baseline. Table \ref{hovernet-result} show the segmentation performances with Hover-Net. For MoNuSeg dataset, it is clear that the segmentation metrics drop with fewer labeled images. For example, with only 10\% labeled images, Dice and AJI reduce by 2.4\% and 3.1\%, respectively. However, by augmenting the 10\% labeled subset, Dice and AJI exceed the fully-supervised baseline by 0.9\% and 1.3\%. For the 20\% and 50\% case, the two metrics obtained by augmented subset are of the same level as using all labeled images. Note that the metrics of 10\% augmented subset are higher than those of 20\% augmented subset, which might be attributed to the indetermination of the diffusion model training and sampling. Interestingly, augmenting the full dataset also helps: Dice increases by 1.3\% and AJI increases by 1.6\% compared with the original full dataset. Therefore, the proposed augmentation method consistently improves segmentation performance of different labeling proportion. For Kumar dataset, by augmenting 10\% labeled subset, AJI increases to a level comparable with that using 100\% labeled images; by augmenting 20\% and 50\% labeled subset, AJIs exceed the fully-supervised baseline. These results demonstrate the effectiveness of the proposed augmentation method that we can achieve the same or higher level segmentation performance of the fully-supervised baseline by augmenting a dataset with a small amount of labeled images.

\subsubsection{Generalization of the proposed data augmentation.}
Moreover, we have similar observations when using PFF-Net as the segmentation model. Table \ref{pffnet-result} shows the segmentation results with PFF-Net. For both MoNuSeg and Kumar datasets, all the four labeling proportions metrics notably improve with synthetic samples. This indicates the generalization of our proposed augmentation method. 

\begin{table}[!t]
\centering
\caption{Effectiveness of the proposed data augmentation method with Hover-Net.}
\begin{tabular}{ p{3cm}|p{1.5cm}|p{1.5cm}|p{1.5cm}|p{1.5cm} }
 \hline
 \multirow{2}{*}{\centering Training data} & \multicolumn{2}{l|}{MoNuSeg} & \multicolumn{2}{l}{Kumar} \\
 \cline{2-5} & Dice & AJI & Dice & AJI \\
 \hline
 10\% labeled & 0.7969 & 0.6344 & 0.8040	& 0.5939  \\
 10\% augmented & 0.8291	& 0.6785 & 0.8049 & 0.6161  \\
 \hline
 20\% labeled & 0.8118 & 0.6501 & 0.8078	& 0.6098  \\
 20\% augmented & 0.8219	& 0.6657 & 0.8192 & 0.6255  \\
 \hline
 50\% labeled & 0.8182 & 0.6603 & 0.8175	& 0.6201  \\
 50\% augmented & 0.8291	& 0.6764 & 0.8158 & 0.6307 \\
 \hline
 100\% labeled & 0.8206 & 0.6652 & 0.8150 & 0.6183  \\
 100\% augmented & 0.8336 & 0.6810 & 0.8210 & 0.6301 \\
 \hline
\end{tabular}
\label{hovernet-result}
\end{table}

\begin{table}[!t]
\centering
\caption{Generalization of the proposed data augmentation method with PFF-Net.}
\begin{tabular}{ p{3cm}|p{1.5cm}|p{1.5cm}|p{1.5cm}|p{1.5cm} }
 \hline
 \multirow{2}{*}{\centering Training data} & \multicolumn{2}{l|}{MoNuSeg} & \multicolumn{2}{l}{Kumar} \\
 \cline{2-5} & Dice & AJI & Dice & AJI \\
 \hline
 10\% labeled & 0.7489	& 0.5290 & 0.7685 & 0.5965   \\
 10\% augmented & 0.7764 & 0.5618 & 0.8051	& 0.6458  \\
 \hline
 20\% labeled & 0.7691	& 0.5629 & 0.7786 & 0.6087  \\
 20\% augmented & 0.7891 & 0.5927 &  0.8019	& 0.6400  \\
 \hline
 50\% labeled & 0.7663	& 0.5661 & 0.7797 & 0.6175  \\
 50\% augmented & 0.7902	& 0.5998 & 	0.8104	& 0.6524\\
 \hline
 100\% labeled & 0.7809 & 0.5708 & 0.8032 & 0.6461  \\
 100\% augmented & 0.7872 & 0.5860 & 0.8125 & 0.6550 \\
 \hline
\end{tabular}
\label{pffnet-result}
\end{table}

\section{Conclusion}
In this paper, we propose a novel diffusion-based data augmentation method for nuclei segmentation in histopathology images. The proposed unconditional nuclei structure synthesis model can generate nuclei structures with realistic nuclei shapes and spatial distribution. The proposed conditional histopathology image synthesis model can generate images of close resemblance to real histopathology images and high diversity. Great alignments between synthetic images and corresponding nuclei structures are ensured by the special design of the conditional diffusion model and classifier-free guidance. By augmenting datasets with a small amount of labeled images, we achieved even better segmentation results than the fully-supervised baseline on some benchmarks. Our work points out the great potential of diffusion models in paired sample synthesis for histopathology images.

\bibliographystyle{splncs04}
\bibliography{ref}

\end{document}